ARTICLE

# Influence of phytohormones on seed germination of *Solanum linnaeanum*

Aram Akram Mohammed, Haidar Anwar Arkwazee, Ayub Karim Mahmood, Hemn Abdalla Mustafa, Hawar Sleman Halshoy, Salam Mahmud Sulaiman, Jalal Hamasalih Ismael & Nawroz Abdul-razzak Tahir

Horticulture Department, College of Agricultural Engineering Sciences, University of Sulaimani, Sulaimani, Kurdistan Region, Iraq.

## Resumen

*Influencia de fitohormonas en la germinación de semillas de Solanum linnaeanum*

El objetivo del estudio fue determinar la capacidad germinativa y crecimiento de plántulas del tomatillo del diablo por inmersión separada en agua, giberelina ($GA_3$), ácido naftilacético (NAA) y ácido salicílico (SA). Los resultados mostraron que el NAA a 50 mgL$^{-1}$ produjo superiores germinación (77,78%), velocidad de germinación (1,43 semillas/tiempo), longitud del hipocotilo (1,01 cm), diámetro del hipocotilo (1,13 mm), número de hojas (2,66) y número de raíces (17,25), seguido de 50 y 100 mgL$^{-1}$ de $GA_3$, particularmente el porcentaje de germinación. La mejor elongación de raíz (5,33 cm) se detectó con 100 mgL$^{-1}$ de SA. Por el contrario, las semillas control y las empapadas en agua mostraron resultados inferiores. Las semillas del tomatillo del diablo pueden germinar con éxito por tratamiento con NAA a 50 mgL$^{-1}$, seguido de $GA_3$ a 50 y 100 mgL$^{-1}$.

**Palabras clave:** Tomatillo del diablo; Ácido 1-naftalenacético (NAA); Ácido gibberélico ($GA_3$); Ácido salicílico (SA).

## Abstract

The aim of this study was to determine the germination ability and seedling growth of the apple of Sodom by soaking in water, gibberellin ($GA_3$), naphthylacetic acid (NAA), and salicylic acid (SA), separately. The findings showed that NAA at 50 mgL$^{-1}$ produced superior germination (77.78%), germination speed (1.43 seeds/time interval), hypocotyl length (1.01 cm), hypocotyl diameter (1.13 mm), leaf number (2.66), and root number (17.25), followed by 50 and 100 mgL$^{-1}$ $GA_3$, particularly in germination percentage. The best root length (5.33 cm) was detected at 100 mgL$^{-1}$ SA. In contrast, control seeds and water-soaked seeds showed inferior results. The seeds of the apple of Sodom can be germinated successfully as a result of treatment with NAA at 50 mgL$^{-1}$, followed by $GA_3$ at 50 and 100 mgL$^{-1}$.

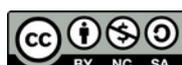

## Introduction

*Solanum linnaeanum* Hepper & P.M.L. Jaeger, which is commonly known as devil's apple or apple of sodom, is a member of the Solanaceae family and native to South Africa (Knapp *et al.* 2013), which has a warmer temperate, sub-tropical and semi-arid condition. It is a perennial bushy shrub up to 1.5 meters tall. The stems are green purplish-brown. The leaves are deeply lobed with prickles on the abaxial and adaxial, and along the stem. The colour of the flowers is purple to white. While, the immature fruits are round mottled green in colour, but turned yellow at maturity (Elabbaraa *et al.* 2014). The fruit of *S. linnaeanum* is poisonous, however secondary metabolites and substances isolated from fruits and other parts of this plant may be used for some health purposes. In this regard, its fruits, leaves, and stems are rich in glycoalkaloids, whereas the whole plant is a good source of solamargine and solasonine, glycoalkaloids that are effective in skin cancer chemotherapy (Ono *et al.* 2006, Gürbüz *et al.* 2015). Also, glycoalkaloids solasodine rhamnosyl glycosides (SRGs), which occur in *S. linnaeanum* and other solanaceous plants, have a role in stimulating apoptosis in a vast assortment of cancer cells (Chauhan 2018). Moreover, Mahomoodally & Ramcharun (2015) referred that *S. linnaeanum* is used traditionally to control diabetes in Mauritius; they found phenols and flavonoids in the methanol extract of the fruits.

In addition to health benefits, *S. linnaeanum* is used for many agricultural purposes. *Solanum linnaeanum* is resistant to several pathogens and can tolerate salinity. Thus, it could be used as a potential genetic source for the hybridization of cultivated eggplants (Weese & Bohs 2010). Liu *et al.* (2015) introduced *Verticillium* Nees wilt resistance successfully into cultivated eggplant from *S. linnaeanum*, and they observed this resistance in the backcrossed population. Apart from this, the leaf extract of *S. linnaeanum* actively mitigated the intensity of Fusarium crown rot (*Fusarium pseudograminearum* O'Donnell & T. Aoki) and Root Rot (FCRR) in tomatoes by 92.3% at 30% (w/v), and significantly improved tomato seed germination as well (Nefzi *et al.* 2018). Similarly, food consumption in *Schistocera gregaria* Forsskål, 1775 fifth instar larva was reduced as a result of the eating lettuce soaked in various extracts of *S. linnaeanum* fruit peel (Zouiten *et al.* 2006). Garzoli *et al.* (2022) detected allelopathic potential in leaf extract of *S. linnaeanum* to inhibit seed germination of some noxious plants. Furthermore, there are reports that *S. linnaeanum* is a compatible rootstock for grafting eggplant as it was obtained from seeds, which is short-term to overcome infestation of pathogens and unfavorable conditions in soil (Villeneuve *et al.* 2016). It is categorized as an invasive plant in some regions of the world (Pyšek *et al.* 2020).

Up to now, no published systematic research has been conducted to reveal the germination capacity of this species. Generally, to enhance germination traits and subsequent growth of the seedlings, many chemical treatments such as plant growth regulators (PGRs) are applied to the seeds of various species (Miransari & Smith 2014). Gibberellin ($GA_3$) is a PGR well-known for its impact on seed germination in aspects like, germination percentage and speed. So, it was applied to *S. nigrum* seed and increased germination percentage to 99% (Rezvani & Fani Yazdi 2013). Salicylic acid (SA) is another PGR that significantly induces seed germination and makes the produced seedlings stronger against many adverse conditions (Moravcová *et al.* 2018). Meanwhile, naphylacetic acid (NAA) is considered as a capable factor to improve seed germination (Li *et al.* 2012). Therefore, the current study was to investigate the effect of different concentrations of $GA_3$, SA, and NAA on seed germination of *Solanum linnaeanum*. The findings of this study inform scientists on how to generate seedlings of this species for the breeding of other solanaceous crops, acquiring secondary metabolites, and other cultural goals.

## Materials and methods

The research was conducted at the College of Agricultural Engineering Sciences, University of Sulaimani, Kurdistan Region-Iraq to examine the germination response of *S. linnaeanum* seeds to different concentrations of $GA_3$, SA, and NAA treatments. The seeds of *S. linnaeanum* were collected from a single fruit produced on a wild plant in Nairobi, Kenya (1º22′24″S 36º51′32″E), on 10th October 2021. The collected seeds were dried at 40 ºC and stored in a paper bag at room temperature (20-27 ºC) for 126 days until the time of sowing.

**Treatment and sowing of the seeds**

The seeds of *S. linnaeanum* were separately soaked in water and different concentrations of (50 and 100 mgL$^{-1}$) of GA$_3$, NAA, and SA for 24 hours, on 14 February 2022 (Li *et al.* 2012). The treated seeds from all treatments were sown in peat medium, which was prepared in a single polystyrene seed tray. While, the seeds were directly sown without any treatment served as control. Three replications were used in the experiment's completely randomized design (CRD), with nine seeds each, for a total of 27 seeds per treatment. The seeds in the tray were placed in a plastic high tunnel. To calculate germination speed, the seeds were checked daily. Also, the average minimum and maximum temperatures inside the high tunnel were between (7.2-33.2 ºC) during the study.

**Harvest, measurement and statistical analysis**

After 10 weeks from sowing, the seedlings were checked to take the measurements. They were taken out of the seed tray cells by a gentle press on the seedlings with a blunt stick via the hole at the bottom of the cells. Afterwards, the seedlings were soaked in water to remove the peat medium and without damaging the roots. Germination percentage, germination speed, leaf number, shoot length, shoot diameter, number of the main roots, and length of the longest main lateral roots were calculated. For calculation of germination percentage and speed, the following formulae were used (Gairola *et al.* 2011).

Germination % = N germinated seeds /Total seeds × 100

Speed of germination= N1/d1+N2/d2+N3/d3+----------

Where, N=number of germinated seeds, d= number of days

The data were tested for homogeneity and analyzed using one-way ANOVA-CRD by XLSTAT software version 2019.2.2. The means were compared according to Duncan's multiple range test (p ≤ 0.05). Principle component (PCA) and Pearson's correlation analysis were applied to the variables to show the degree of proximity between them.

## Result and discussion

The data are shown in the figure 1 refers to the germination percentage and germination speed of *S. linnaeanum* seeds after exposing to water soaking and soaking in GA$_3$, NAA and SA for 24h. The findings proved that 50 and 100 mgL$^{-1}$ of NAA and GA$_3$ significantly increased germination when compared to control seeds (Fig. 1A). Accordingly, the highest numbers of germinated seeds (77.78%) were found among the ones soaked in 50 mgL$^{-1}$ NAA followed by 50 mgL$^{-1}$ GA$_3$ and 100 mgL$^{-1}$ NAA (74.07%). Inversely, control seeds gave the lowest germination (48.15%). It is noteworthy to mention that the germination pattern in *S. linnaeanum* was epigeal. Despite germination percentage, the treatments were effective to increase germination speed, and different germination speeds were found depending on the treatment (Fig. 1B). Soaking the seeds in 50 mgL$^{-1}$ NAA hastened germination to the maximum level (1.43 seed/time interval), but germination was the slowest in control seeds (0.79 seed/time interval), the seeds were soaked in water (0.81 seed/time interval), and those were treated with 100 mgL$^{-1}$ SA (0.83 seed/time interval). The seeds that were soaked in NAA and GA$_3$ started to germinate 24 days after sowing, but the control seeds and those that were soaked in the water started to germinate 30 days after sowing. The germination process is under many endogenous and exogenous factors among them growth regulators have a decisive influence on this process (Aticia *et al.* 2003). In the current study, application of NAA and GA$_3$ was outstanding to improve seed germination of *S. linnaeanum*. Regarding the concentration and plant species, NAA has been emphasized as an inducible agent in the germination of the seeds. Shin *et al.* (2011) observed that NAA increased germination of Calanthe hybrids seed at 0.1 mgL$^{-1}$. Also, Maku *et al.* (2014) stated that treatment of *Tetrapleura tetraptera* (Schumach. & Thonn.) Taub. seeds with NAA induced germination progressively, and they attributed this result to the effect of NAA on the seed coat. Apart from these, ethylene biosynthesis might be elevated because of NAA treatment, hence ethylene may enhance germination as well (Li & Yuan 2008, Kępczyński & Van Staden 2012). On the other hand, gibberellin is frequently used to maximise germination percentage and hasten germination in most species. It activates the enzymes needed to mobilize and metabolite the stored foods in the endosperm, weakens the constraints on embryo due to endosperm or seed coat, and elongates the cells of the embryo (Hartmann *et al.* 2011). Besides, the inhibitory consequences of abscisic acid (ABA) on the germination of the seeds could be nullified by GA$_3$ application. It is also possible exogenous applied GA$_3$ has potential to convert into endogenous GA$_1$ which also induces germination (Chen *et al.* 2008).

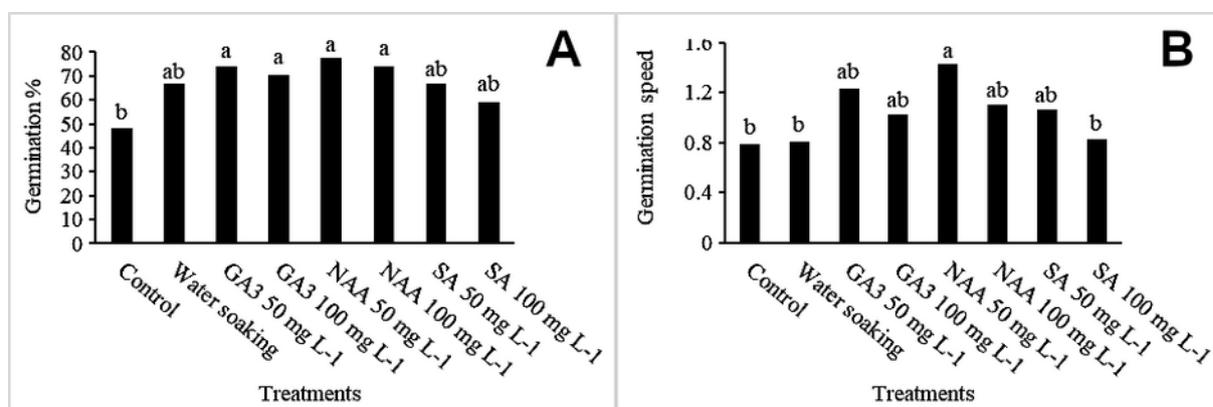

**Figura 1.** Efecto del remojo en agua y de diferentes concentraciones de giberelina (GA$_3$), ácido naftilacético (NAA) y ácido salicílico (SA) sobre semillas de *S. linnaeanum*. **A:** porcentaje de germinación; **B:** velocidad de germinación. Las barras con la misma letra no difieren significativamente según la nueva prueba de rango múltiple de Duncan (p≥0,05).

**Figure 1.** Effect of water soaking and different concentrations of gibberellin (GA$_3$), naphthylacetic acid (NAA), and salicylic acid (SA) on *S. linnaeanum* seeds. **A:** germination percentage; **B:** germination speed. Bars with the same letter do not differ significantly according to the Duncan's new multiple range test (p≥0.05).

The data in the figure 2 explained how the hypocotyl length, hypocotyl diameter, and leaf number of germinated seeds of *S. linnaeanum* responded differently to the treatments. The treated seeds had different hypocotyl lengths compared to the control seeds after being soaked in water and exposed to 50 mgL$^{-1}$ NAA (Fig. 2A). The longest hypocotyl (1.01 cm) was achieved at 50 mgL$^{-1}$ NAA, while the shortest hypocotyl (0.7

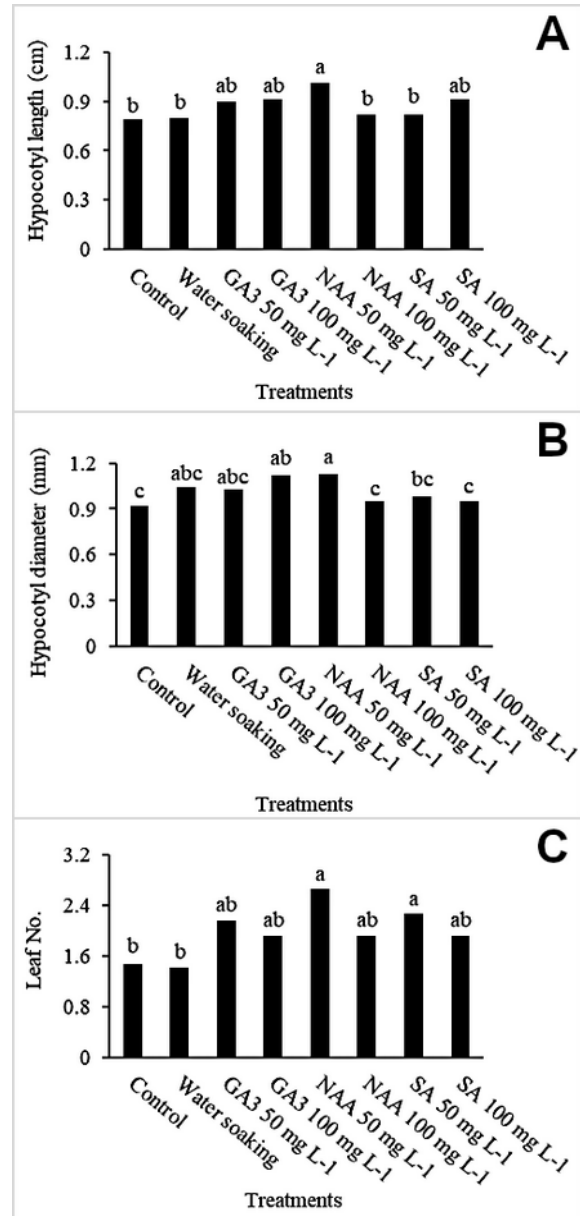

**Figura 2.** Efecto del remojo en agua y diferentes concentraciones de giberelina (GA$_3$), ácido naftilacético (NAA) y ácido salicílico (SA) sobre semillas de *S. linnaeanum*. **A:** longitud del hipocotilo (cm); **B:** el diámetro del hipocotilo; **C:** número de hojas. Las barras con la misma letra no difieren significativamente según la nueva prueba de rango múltiple de Duncan (p≥0,05)

**Figure 2.** Effect of water soaking and different concentrations of gibberellin (GA$_3$), naphthylacetic acid (NAA) and salicylic acid (SA) on *S. linnaeanum* seeds. **A:** hypocotyl length (cm); **B:** hypocotyl diameter; **C:** leaf number. Bars with the same letter do not differ significantly according to the Duncan's new multiple range test (p≥0.05).

cm) was recorded in germinated control seeds and in water-soaked seeds (0.8 cm). However, by increasing NAA concentrations from 50 to 100 mgL$^{-1}$, hypocotyl length was decreased, and 50 mgL$^{-1}$ SA was insufficient to significantly lengthen the hypocotyl. Meanwhile, the treatments especially 50 mgL$^{-1}$ NAA and 100 mgL$^{-1}$

GA$_3$ caused the hypocotyl diameter to increase significantly when they were compared to control group (Fig. 2B). The hypocotyls with the largest diameter (1.13 mm) were measured at 50 mgL$^{-1}$ NAA treatment, followed by those soaked in 100 mgL$^{-1}$ GA$_3$ (1.12 mm). The control seeds produced the thinnest hypocotyl (0.92 mm), while the seeds soaked in 100 mgL$^{-1}$ NAA and SA produced seedlings with the same diameter (0.95 mm). Similarly, soaking the seeds in different treatments results in various leaf numbers, without cotyledons (Fig. 2C). The best number of leaves (2.66 and 2.27) were obtained at 50 mgL$^{-1}$ NAA and SA, respectively. While, the lowest leaf number was found in the seedlings from the seeds were soaked in water (1.41) and control seeds (1.47). On the whole, the best for hypocotyl growth was measured at 50 mgL$^{-1}$ NAA. These results reveal the fact that NAA at 50 mgL$^{-1}$ promoted earlier germination (Fig. 1B), and by doing so the germinated seeds had more time for hypocotyl growth and developing more leaves. Moreover, auxins play an essential role in hypocotyl growth. It was found that auxin at high concentrations and in light conditions stimulated hypocotyl growth (Muday *et al.* 2012). Further, Anwar *et al.* (2020) discovered in seeds of cucumber that GA$_3$ demonstrated the maximal hypocotyl diameter. Hypocotyl growth is favorably maintained by GAs by degrading DELLA proteins, which have a detrimental impact on growth (Hauvermale *et al.* 2012). There are reports that SA at low concentration increased leaf number as well, as was observed in this study (Yusuf *et al.* 2013).

The application of the growth regulators in the present study linked to a better root length and root number in germinated seeds of *S. linnaeanum* (Fig. 3). The longest roots (5.33 and 5.27 cm) were measured in the seedlings germinated from the soaked seed in 100 mgL$^{-1}$ SA and 50 mgL$^{-1}$ NAA, respectively (Fig. 3A). Oppositely, soaking the seeds in water did not encourage root length significantly and they gave the shortest roots (4.04 cm). Likewise, a variable main root number was detected according to treated and untreated seeds with 50 and 100 mgL$^{-1}$ of the three growth regulators (Fig. 3B). The highest root number (17.25) was created by NAA at 100 mgL$^{-1}$, and it varied with control and water-soaked seeds. Control seeds generated the lowest root number (10.27), accompanied by water-soaked seeds (12.41). Furthermore, root number was better in the seedlings germinated from the seeds soaked in 100 mgL$^{-1}$ GA$_3$, 50 mgL$^{-1}$ NAA, and 50 mgL$^{-1}$ SA than the control seedlings. The results of root trait measurements could be due to the effect of growth regulators on faster germination, which allowed the produced leaves to be exposed to the sun sooner and conduct photosynthesis for a longer period of time, producing more photosynthates for better root development. In this context, Muralidhara *et al.* (2016) revealed that rapid germination is an opportunity for the development of a good root system and to augment photosynthesis through early sunlight exposure to the leaves. Additionally, growth regulators might enhance root growth and formation by enhancing cell division and elongation. Auxins, including NAA, are vital to promoting cell proliferation by acting on cyclin-dependent kinases (CDKs); adding NAA diminished transcription of CDK inhibitors KRP1 and KRP2, and thereby formation of lateral roots was simultaneously elevated (Himanen *et al.* 2002). The superior effect of 100 mgL$^{-1}$ SA on root length in the present study could be attributed to the stimulatory action of SA to regulate some physiological and biochemical processes required for better root growth. Soliman *et al.* (2016) stated that SA in the apical meristem of seedling roots upgrades cell division, cell differentiation, and cell elongation; these collectively lead to longer roots.

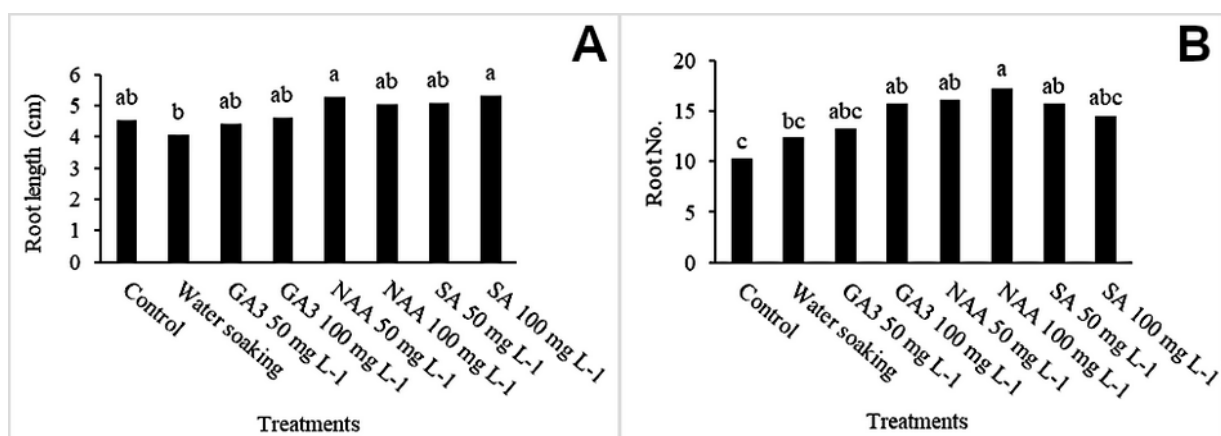

**Figura 3.** Efecto del remojo en agua y diferentes concentraciones de giberelina (GA$_3$), ácido naftilacético (NAA) y ácido salicílico (SA) en semillas de *S. linnaeanum*. **A:** longitud de la raíz (cm); **B:** número de raíces. Las barras con la misma letra no difieren significativamente según la nueva prueba de rango múltiple de Duncan (p≥ 0,05).

**Figure 3.** Effect of water soaking and different concentrations of gibberellin (GA$_3$), naphthylacetic acid (NAA), and salicylic acid (SA) on *S. linnaeanum* seeds. **A:** root length (cm); **B:** root number. Bars with the same letter do not differ significantly according to the Duncan's new multiple range test (p≥ 0.05).

**Interrelationship analysis of the variables**

PCA were undertaken to illustrate the degree of proximity between variables (Fig. 4). PCA1 and PCA2 accounted for 82.39% of the total variation and manifested as two axes of variances. The horizontal axis (PCA1) explains 63.35% of variations, whereas the vertical axis (PCA2) explains 19.04% of variations. PCA1 has a strong and positive correlation with the following characteristics: NAA 50, G, GS, HL, RN, HD, and LN. In addition, high associations were identified between G, GS, HL, RN, HD, and LN characteristics and NAA 50 treatment, demonstrating that NAA at a concentration of 50 mgL$^{-1}$ induces these traits. Similarly, PCA2 was favorably associated with RL and adversely associated with HD. The water soaking treatment was positioned on the negative side of PCA1 and PCA2 and was adversely linked with RL, RN, and LN, whereas the control and SA 100 were inversely associated with G, GS, HL, and HD. In addition, PCA (Fig. 4) and Pearson's correlation analysis (Fig. 5) confirmed positive correlations between G% and GS (p = 0.013), GS and HL (p = 0.05), GS and LN (p=0.003), and HL and LN (p = 0.029).

## Conclusion

This research revealed that soaking *S. linnaeanum* seeds in 50 mgL$^{-1}$ NAA for 24 h before sowing in peat medium was effective in encouraging germination and the other studied parameters after germination. Also, GA$_3$ at 50 and 100 mgL$^{-1}$ gave the highest germination percentage, and SA showed the best root growth similar to NAA. NAA at 100

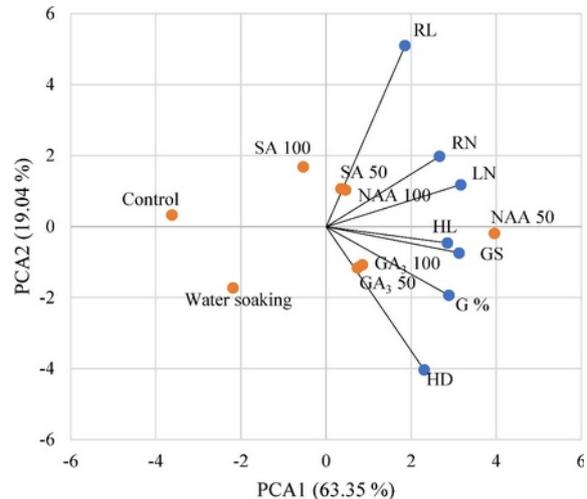

**Figura 4.** Relación y posición de las variables independientes y dependientes. GA$_3$ 50 y GA$_3$ 100: giberelina a 100 y 50 mgL$^{-1}$; NAA 50 y NAA 100: ácido naftilacético a 50 y 100 mgL$^{-1}$; SA 50 y SA 100: ácido salicílico a 50 y 100 mgL$^{-1}$. G%: porcentaje de germinación; GS: velocidad de germinación; HL: longitud del hipocotilo; HD: diámetro del hipocotilo; LN: número de hojas; RL: longitud de las raíces; RN: número de raíces.
**Figure 4.** Relationship and position of the independent and dependent variables. GA$_3$ 50 and GA$_3$ 100: gibberellin at 100 and 50 mgL$^{-1}$; NAA 50 and NAA 100: naphthylacetic acid at 50 and 100 mgL$^{-1}$; SA 50 and SA 100: salicylic acid at 50 and 100 mgL$^{-1}$. G%: germination percentage; GS: germination speed; HL: hypocotyl length; HD: hypocotyl diameter; LN: leaf number; RL: root length; RN: root number.

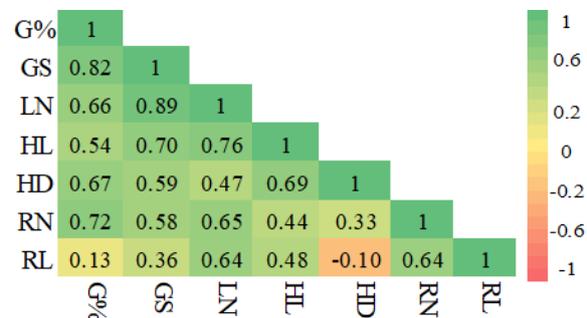

**Figura 5.** Análisis de correlación de Pearson de los caracteres estudiados (p=0,05). G%: porcentaje de germinación; GS: velocidad de germinación; HL: longitud del hipocotilo; HD: diámetro del hipocotilo; LN: número de hojas; RL: longitud de la raíz; RN: número de raíces.
**Figure 5.** Pearson's correlation analysis of the studied traits, (p=0.05). G%: germination percentage, GS: germination speed; HL: hypocotyl length; HD: hypocotyl diameter; LN: leaf number, RL: root length; RN: root number.

mgL$^{-1}$ and SA at 50 and 100 mgL$^{-1}$ reduced hypocotyl growth. Meanwhile, the lowest developments were obtained from control seeds and, to a lesser extent, water-soaked seeds. Despite this, more research is needed to determine the degree of dormancy, if it occurs, and whether it is a polymorphic seed species that allows understanding of its invasiveness in some regions and survival in extreme conditions.